# FIFO anomaly is unbounded

Péter Fornai
Eötvös Loránd University
Department of Computer Algebra
H-1117, Budapest, Hungary
Pázmány sétány 1/C
email: peter.fornai@inf.elte.hu

Antal Iványi
Eötvös Loránd University
Department of Computer Algebra
H-1117, Budapest, Hungary
Pázmány sétány 1/C
email: tony@inf.elte.hu

**Abstract.** Virtual memory of computers is usually implemented by demand paging. For some page replacement algorithms the number of page faults may increase as the number of page frames increases. Belady, Nelson and Shedler [5] constructed reference strings for which page replacement algorithm FIFO [10, 13, 36, 40] produces near twice more page faults in a larger memory than in a smaller one. They formulated the conjecture that 2 is a general bound. We prove that this ratio can be arbitrarily large.

## 1 Introduction

Let us consider a computer with two-level virtual memory [12, 26, 38]. Physical and secondary memory are divided into equal-sized blocks called frames. The processes reside in secondary memory and are broken into fixed-sized blocks called pages (the sizes of the frames and pages are the same). The run of the processes is described by reference strings.

According to the pure demand paging the execution of a process starts with empty physical memory (in the following simply *memory*) and a page is brought into the memory only when it is necessary. If a new page is demanded when the memory is full, the page replacement algorithm [38] chooses a page being in memory to be replaced by the demanded one.

We use the formal model proposed by P. Denning [12]. Let $m$, $M$, $n$, and $p$ be positive integers ($1 \leq m \leq M \leq n < \infty$), $k$ a nonnegative integer, $A = \{a_1, a_2, \ldots, a_n\}$ a finite alphabet, $A^k$ the set of k-length words and $A^*$ be the set of finite words over $A$, where $n$ is the number of pages, $m$ is the number of frames in the small and $M$ is the number of frames in the large memory, $A$ is the set of pages.

Page replacement algorithms [38] are handled as automata having a memory of size $m$ (or $M$), set of input signals $X = A$, set of output signals $Y = A \cup \{\epsilon\}$ and processing the sequence of input symbols $R = (r_1, r_2, \ldots, r_p)$ or $R = (r_1, r_2, \ldots)$. Memory state $S_t$ ($t = 1, 2, \ldots$) is defined as a set of symbols stored in memory at the given moment $t$ (after the processing of reference $r_t$). Any page replacement algorithm starts with an empty memory that is $S_0 = \{\}$. A concrete page replacement algorithm P is defined as a triple $P = (Q_P, q_0, g_P)$, where $Q_P$ is the set of control states, $q_0 \in Q_P$ is the initial control state and $g_P$ is the transition function determining the new memory state $S'$, new control state $q'$ and output symbol $y$ using the old memory state $S$, old control state $q$ and input symbol $x$.

We consider three page replacement algorithms: FIFO (First In First Out) [10, 13, 36, 40], LRU (Least Recently Used) [8, 10, 19, 26, 37] and MIN (Minimal) [1, 2, 5, 28, 29, 32, 34].

FIFO is defined by $q_0 = ()$ and

$$g_{\mathrm{FIFO}}(S, q, x) = \begin{cases} (S, q, \epsilon), & \text{if } x \in S, \\ (S \cup \{x\}, q', \epsilon), & \text{if } x \notin S \text{ and } |S| < m, \\ (S \cap \{y_1\} \cup \{x\}, q'', y_1), & \text{if } x \notin S \text{ and } |S| = m, \end{cases}$$

where $q = (y_1, y_2, \ldots, y_k)$, $q' = (y_1, y_2, \ldots, y_k, x)$, and $q'' = (y_2, y_3, \ldots, y_m, x)$.

LRU replaces the least recently used page of the memory, MIN replaces the page having maximal distance up to its next occurrence in the reference string.

The number of page faults (number of changes of the memory states) is denoted by $f_P(R, m)$. If $M > m$ and $f_P(R, M) > f_P(R, m)$, then we have **an anomaly** and the ratio $f_P(R, M)/f_P(R, m)$ is called **anomaly ratio**. The first anomaly was observed in the practice [4, 5, 34] and is called Belady's anomaly [28, 35]. Later other examples of unexpected events were described [6, 9, 14, 17, 18, 20, 21, 22, 23, 25, 27, 31, 33].

Stack algorithms [28, 30, 39] do not suffer from anomaly.

Mihnovskiy and Shor [24] and Gecsei et al. [15] proved independently that MIN guarantees the minimal number of page faults at a fixed size of the memory.

Arató [3] and Benczúr et al. [7] proved that LRU is optimal in statistical sense among the algorithms having no concrete information on the continuation of the reference string.

A possible measure of the efficiency $E_P(R, m)$ of P is called **paging rate** and is defined by $f_P(R, m)/p$ for finite $R = (r_1, r_2, \ldots, r_p)$ and by

$$E_P(R, m) = \liminf_{k \to \infty} \frac{f_P(R_k, m)}{k}$$

for an infinite reference string R, where $R_k = (r_1, r_2, \ldots, r_k)$.

Let $1 \leq m < n$ and $C = (1, 2, \ldots, n)^*$ be an infinite cyclical reference string. Then the paging rates of FIFO and MIN are given in [16] $E_{FIFO}(C, m) = 1$ and $E_{MIN}(C, m) = (n - m)/(n - 1)$.

## 2 Classical example and results

If we execute the reference string $R = (1,2,3,4,1,2,5,1,2,3,4,5)$ [5] having 3 frames in the memory (for the simplicity natural numbers are used to denote the pages), then FIFO results the control state sequence $q_0 = ()$, $q_1 = (1)$, $q_2 = (1, 2)$, $q_3 = (1, 2, 3)$, $q_4 = (2, 3, 4)$, $q_5 = (3, 4, 1)$, $q_6 = (4, 1, 2)$, $q_7 = (1, 2, 5)$, $q_8 = (1, 2, 5)$, $q_9 = (1, 2, 5)$, $q_{10} = (2, 5, 3)$, $q_{11} = (5, 3, 4)$, $q_{12} = (5, 3, 4)$ and 9 page faults occurred.

The execution of R using 4 frames will end in the control state $(2,3,4,5)$ and 10 page faults occurred, so $f_{FIFO}(R, M)/f_{FIFO}(R, m) = 10/9$.

Belady, Nelson and Shedler [5] gave a necessary and sufficient condition for the existence of anomaly and constructed reference strings resulting anomaly ratio which is close to 2.

**Theorem 1** [5] *Let* m *and* M *be positive integers* $(1 \leq m < M)$ *and let A be a finite alphabet of cardinality at least* $M + 1$. *Then there exists a reference string* $R \in A^*$ *which produces an anomaly if and only if* $M < 2m - 1$.

**Theorem 2** [5] *If* m *and* M *are positive integers satisfying the relations* $m < M < 2m - 1$, *then for sufficiently large* p *there exists a reference string* $R = (r_1, r_2, \ldots, r_p)$ *resulting anomaly ratio* $f_{FIFO}(R, M)/f_{FIFO}(R, m)$ *arbitrarily close to 2.*

Belady, Nelson and Shedler [5] formulated the following conjecture.

**Conjecture** [5] *If $m$ and $M$ are positive integers and $R$ is a reference string, then $f_{\text{FIFO}}(R, M)/f_{\text{FIFO}}(R, m) \leq 2$.*

## 3 Disprove of the conjecture

Let $m = 5$, $M = 6$, $n = 7$, $k \geq 1$ and $R = UV^k$, where $U = (1,2,3,4,5,6,7,1,2,$
$4,5,6,7,3,1,2,4,5,7,3,6,2,1,4,7,3,6,2,5,7,3,6,2,5)$ and $V = (1,2,3,4,5,6,7)^3$.

Now execution of $U$ using $m$ frames results the control state $(7,3,6,2,5)$ and 29 page faults. After this each execution of $V$ results 7 new faults and the same control state. Execution of $U$ using $M$ frames results the control state $(2,3,4,5,6,7)$ and 14 page faults. After this each execution of $V$ results 21 new page fault and the same control state.

Choosing $k = 7$ we get an anomaly ratio $(14+7\times21)/(29+7\times7) = 161/78 > 2$. If the number of repetitions of $V$ grows, then the anomaly ratio tends to 3.

## 4 Anomaly is unbounded

We start with a well-known assertion of the number theory. Let $[a]_n$ be the equivalence class modulo $n$ containing the number $a$ and let $\mathbf{Z}_n$ be the set of all equivalence classes modulo $n$:

$$[a]_n = \{a + kn : k \in \mathbf{Z}\} \text{ and } \mathbf{Z}_n = \{[a]_n : 0 \leq a \leq n-1\}.$$

**Lemma 3** [16] *If the set $c_1, c_2, \ldots, c_n$ is a complete residue system $\mathbf{Z}_n \pmod{n}$, the greatest common divisor of $a$ and $n$ equals 1, then the set $\{ac_1 + d, ac_2 + d, \ldots, ac_n + d\}$ is also a complete residue system $\pmod{n}$ for arbitrary integer $d$.*

Let $a \pmod{b}$ denote the smallest positive representative of the residue class of numbers congruent with $a \pmod{b}$.

**Lemma 4** *Let $n$ be an odd positive integer. Then the elements of the sequence $W = (w_1, w_2, \ldots, w_n) = \bigl(1 \pmod{n}, 1 + (n-1)/2 \pmod{n}, 1 + 2(n-1)/2 \pmod{n}, \ldots, 1 + (n-1)(n-1)/2 \pmod{n}\bigr)$ have the following properties.*
*a) They form a complete system of residues $\pmod{n}$.*

b) If $W^k = s_1, s_2, \ldots, s_{k \times n}$ is the concatenation of $k$ copies of $W$, then any $n$ neighbouring elements of $W^k$ form a complete residue system $(\mathrm{mod}\ n)$.
c) If $j \geq 1$, then $s_{j+n+1} \equiv s_{j+1}\ (\mathrm{mod}\ n)$.

**Proof.** It is sufficient to use Lemma 3 and some elementary calculations. □

**Lemma 5** *Let $n$ be an odd number $(n \geq 5)$, $M = n - 1$ and $m = n - 2$. If the small memory starts with control state $(w_3, w_4, \ldots, w_n)$ and the large memory starts with control state $(2, 3, \ldots, n)$, then the reference string $R = (1, 2, \ldots, n)^{n-1)/2}$ results n page faults in the small memory, $n(n-1)/2$ page faults in the large memory and the initial control states in both memories.*

**Proof.** This is a consequence of Lemma 4. □

**Lemma 6** *Let $1 \leq m < M < n$ and $b_1, b_2, \ldots, b_m$ be arbitrary different elements of $A = \{1, 2, \ldots, M+1\}$. Then algorithm ANOMALY constructs a reference string resulting $q = (b_1, b_2, \ldots, b_m)$ in the small memory of size $m$ while the pages are loaded into the large memory of size $M$ in cyclical order $1, 2, \ldots, n$.*

The next algorithm is written in PASCAL-like pseudocode [11].

CONSTRUCTION ALGORITHM ANOMALY$(m, M, q, U)$.
*Input*: $m$ size of the small memory; $M$ size of the large memory; $q = (b_1, b_2, \ldots, b_m)$ the required control state of the small memory.
*Output*: $U = (u_1, u_2, \ldots)$ reference string.
*Working variables*: $S$ actual state of the small memory; $t$ index of the next reference; $A = \{1, 2, \ldots, M, M+1\}$ the set of pages.

```
 1. for k ← 1 to M
 2.     do u_k ← k
 3. t ← M + 1
 4. i ← 0
 5. while i < m do
 6.     if b_{i+1} ∈ S
 7.         then while b_{i+1} ∈ S do
 8.             u_t ← min{k|k ∈ A and k ∉ S}
 9.             t ← t + 1
10.         while b_1 ∈ S(m) do
11.             u_k ← min{k|k ∈ A and k ∉ S and k ≠ b_{i+1}}
12.             t ← t + 1
```

```
13.                     j ← 1
14.                     while  j < i + 1 do
15.                         u_t ← b_j
16.                         t ← t + 1
17.                         j ← j + 1
18.           u_t ← b_{i+1}
19.           t ← t + 1
20.           i ← i + 1
```

Instead of a long proof of correctness of ANOMALY we explain how this algorithm generates $U$ of the example used to disprove the conjecture. The input of the algorithm is $m = 5$, $M = 6$ and $q = (7,3,6,2,5)$. Lines 1–2 result the references 1, 2, 3, 4, 5, 6 and control state $q = (2,3,4,5,6)$ in the small memory. Now line 3 results $t = 7$, line 4 gives $i = 0$ and we start the while cycle in line 5. Since $b_1 = 7$ is missing from the small memory, we add $u_7 = 7$ to $U$ (lines 18–20) changing the control state to $q = (3, 4, 5, 6, 7)$, increment $t$ and $i$ and return to line 5 where we observe that $b_2 = 3$ is in the memory. Therefore we continue with the while cycle in line 7 and add the minimal missing page to the reference string resulting control state (4,5,6,7,1). Since page 1 replaced $b_2$, we go to the next cycle in lines 10–12 and add the minimal missing pages differing from $b_2 = 3$ (that is 2, 4, 5, and 6) to $U$ changing the control state to (1,2,4,5,6). Now this while cycle ends and using lines 13–17 we add 7 then using lines 18–20 add 3 to $U$ ending in control state (4,5,6,7,3).

Now we return to line 5 and using lines 7-9 add 1 to $U$, then using lines 10–12 add 2, 4 and 5 to $U$ changing the control state to (3,1,2,4,5). Now using lines 14–18 we add 7 and 3, then using lines 19–21 add 6 to $U$ changing the control state to (4,5,7,3,6). Now lines 5–6 send us to lines 14–18 resulting reference $u_{22} = 2$ and control state (5,7,3,6,2).

We observe in lines 5–6 that $b_5$ is in the memory therefore we add 1 to $U$ and so remove 2 from the memory. Now we continue $U$ with 4 resulting the control state (3,6,2,1,4). Finally lines 14–18 add 7, 3, 6 and 2 and lines 19–21 add 5 to $U$ implying the required control state.

**Lemma 7** *Let $n$ be an odd positive integer ($n \geq 5$) and let $M = n - 1$, $m = n - 2$. Then there exists a reference string $R$ resulting anomaly ratio arbitrarily close to $(n - 1)/2$.*

**Proof.** This assertion is a consequence of Lemma 5 and Lemma 6. □

**Main Theorem** *For any large number $L$ there exist parameters $m$, $M$, and $R$ such that the anomaly ratio $f_{FIFO}(R, M)/f_{FIFO}(R, m) > L$.*

**Proof.** Let $n$ be an odd integer satisfying $n > 2L + 1$ (and $n \geq 5$). Then the parameters $m, M$ and $R$ in Lemma 5 result an anomaly ratio greater than $L$. □

## 5 Summary

If the memory parameters $m$ and $M$ are fixed, then the anomaly is bounded.

We suppose that the maximal anomaly occurs at cyclical reference strings, similar memory sizes, FIFO-like replacement in the large memory and MIN-like replacement in the small memory. If so then for fixed odd $n$ our construction results the maximal possible anomaly ratio. We remark that there is a similar construction for even $n$ resulting anomaly ratio arbitrarily close to $(n-2)/2$.

## Acknowledgements

The authors are indebted to Mihály Molnár, former student of Eötvös Loránd University who using genetic algorithms discovered the first reference string resulting an anomaly ratio greater than 2. We thank the useful comments of Professor of Eötvös Loránd University András Benczúr and of the unknown referee.

The research was supported by the project TÁMOP-4.2.1/B-09/1/KMR-2010-003 of Eötvös Loránd University.

## References

[1] S. Albers, On generalized connection caching, in: *ACM Symposium on Parallel Algorithms and Architectures* (Bar Harbor, 2000), *Theory Comput. Syst.*, **35,** (3) (2002) 251–267. ⇒81

[2] S. Albers, L. M. Favrholdt, O. Giel, On paging with locality of reference, *J. Comput. System Sci.*, **70,** 2 (2005) 145–175. ⇒81

[3] M. Arató, A note on optimal performance of page storage, *Acta Cybernet.* (Szeged), **3,** 1 (1976/77) 25–30. ⇒82

[4] L. A. Belady, A study of replacement algorithms for a virtual storage computer, *IBM Syst. J.*, **5,** 2 (1965) 78–101. ⇒81

[5] L. A. Belady, R. A. Nelson, G. S. Shedler, An anomaly in space-time characteristics of certain programs running in paging machine, *Comm. ACM*, **12,** 1 (1969) 349–353. ⇒80, 81, 82, 83

[6] L. A. Belady, F. P. Palermo, On-line measurement of paging behavior by the multivalued MIN algorithm, *IBM J. Res. Develop.*, **18** (1974) 2–19. ⇒81

[7] A. Benczúr, A. Krámli, J. Pergel, On the Bayesian approach to optimal performance of page hierarchies, *Acta Cybernet.* (Szeged), **3,** 2 (1976/77) 78–79. ⇒82

[8] J. Boyar, L. M. Favrholdt, K. S. Larsen, The relative worst-order ratio applied to paging, *J. Comput. System Sci.*, **73,** 5 (2007) 818–843. ⇒81

[9] M. Caminada, L. Amgoud, On the evaluation of argumentation formalisms, *Artificial Intelligence*, **171,** 5–6 (2007) 286–310. ⇒81

[10] M. Chrobak, J. Noga, LRU is better than FIFO, *Algorithmica*, **23,** 2 (1999) 180–185. ⇒80, 81

[11] T. H. Cormen, C. E. Leiserson, R. L. Rivest, C. Stein, *Introduction to algorithms* (Second edition), The MIT Press/McGraw-Hill Book Company, Cambridge/Boston, 2009. ⇒84

[12] P. Denning, Virtual memory. *Comput. Surveys*, **2,** 3 (1980) 153–189. ⇒80, 81

[13] L. Epstein, Y. Kleiman, J. Sgall, R. van Stee, Paging with connections: FIFO strikes again, *Theoret. Comput. Sci.*, **377,** 1–3 (2007) 55–64. ⇒80, 81

[14] M. A. Franklin, G. S. Graham, R. K. Gupta, Anomalies with variable partition paging algorithms, *Comm. ACM*, **21,** 3 (1978) 232–236. ⇒81

[15] J. Gecsei, D. R. Slutz, I. L. Traiger, Evaluation techniques for storage hierarchies, *IBM Syst. J.*, **9,** 2 (1970) 78–117. ⇒81

[16] A. M. Iványi and L. N. Korolev, Classification of paging algorithms, in: *Problems of Interpretation of Experiments* (eds. A. N. Tihonov, P. N. Zaikin and V. Y. Galkin), Moscow State University, Moscow, 1975, pp. 105–125. ⇒82, 83

[17] A. M. Iványi, On dumpling-eating giants, in: *Finite and Infinite Sets* (eds. A. Hajnal, L. Lovász and V. T. Sós), North-Holland, Amsterdam, 1984, pp. 379–390. ⇒81

[18] A. M. Iványi, R. L. Smelyanskiy, *Elements of theoretical programming* (in Russian), Moscow State University, Moscow, 1980. ⇒81

[19] Ya. A. Kogan, A class of hierarchical paging algorithms, *MTA Számitástechn. Automat. Kutató Int. Tanulmányok* (3rd Visegrád Winter School *Theory of Operating Systems*, Visegrád, 1977), No. 69 (1977) 7–13. ⇒81

[20] S. Kolahi, Dependency-preserving normalization of relational and XML data, *J. Comput. System Sci.*, **73,** 4 (2007) 636–647. ⇒81

[21] R. Korf, Linear-time disk-based implicit graph search, *J. ACM*, **55,** 6 (2008) Art. 26, pp. 40. ⇒81

[22] K. V. Malkov, D. V. Tunitskii, On an extremal problem of adaptive machine learning that is connected with the detection of anomalies (in Russian), *Avtomat. i Telemekh.*, 2008, no. 6, 41–52; translation in *Autom. Remote Control*, **69,** 6 (2008) 942–952. ⇒81

[23] B. Mans, C. Roucairol, Performances of parallel branch and bound algorithms with best-first search, *Discrete Appl. Math.*, **66,** 1 (1996), 57–74. ⇒81

[24] S. D. Mihnovskiy, N. Z. Shor, Estimation of the number of page faults in paged virtual memory (in Russian), *Kibernetika*, **1,** 5 (1965) 18–20. ⇒81

[25] D. A. Naumann, Observational purity and encapsulation, *Theoret. Comput. Sci.*, **376,** 3 (2007) 205–224. ⇒81

[26] H. Raquibul, R. Sohel, DesynchLRU: An efficient page replacement algorithm with desynchronized cache and RAM, in: *Abstracts of Int. Conf. on Applied Informatics* (Eger, January 27–30, 2010). ⇒80, 81

[27] S. Roosta, *Parallel processing and parallel algorithms*, Springer-Verlag, New York, 1999. ⇒81

[28] A. Silberschatz, P. Galvin and G. Gagne, *Applied operating system Concepts*, John Wiley and Sons, New York, 2000. ⇒81

[29] D. Sleator, R. E. Tarjan, Amortized efficiency of list update and paging rules, *Comm. ACM*, **28,** 2 (1985) 202–208. ⇒81

[30] A. J. Smith, Analysis of the optimal, look-ahead demand paging algorithms, *SIAM J. Comput.*, **5,** (4) (1976) 743–757. ⇒81

[31] A. M. Sokolov, Vector representations for efficient comparison and search for similar strings (in Russian), *Kibernet. Sistem. Anal.*, **43,** 4 (2007), 18–38, 189; translation in *Cybernet. Systems Anal.*, **43,** 4 (2007) 484–498. ⇒81

[32] H.-G. Stork, On the paging-complexity of periodic arrangements, *Theoret. Comput. Sci.*, **4,** 2 (1977) 171–197. ⇒81

[33] N. B. Waite, A real-time system-adapted anomaly detector, *Inform. Sci.*, **115,** 1–4 (1999) 221–259. ⇒81

[34] Wikipedia, Belady's algorithm, http://en.wikipedia.org/wiki/Belady%27s_Min#Belady.27s_Algorithm, 2010. ⇒81

[35] Wikipedia, Belady's anomaly, http://en.wikipedia.org/wiki/Belady%27s_anomaly, 2010. ⇒81

[36] Wikipedia, FIFO page replacement algorithm, http://en.wikipedia.org/wiki/FIFO, 2010. ⇒80, 81

[37] Wikipedia, LRU page replacement algorithm, http://en.wikipedia.org/wiki/LRU, 2010. ⇒81

[38] Wikipedia, Page replacement algorithm, http://en.wikipedia.org/wiki/Page_replacement_algorithm, 2010. ⇒80, 81

[39] C. Wood, Christopher, E. B. Fernandez, T. Lang, Minimization of demand paging for the LRU stack model of program behavior. *Inform. Process. Lett.*, **16,** 3 (1983) 99–104. ⇒81

[40] N. E. Young, On-line paging against adversarially biased random inputs, (*Ninth Annual ACM-SIAM Symposium on Discrete Algorithms*, San Francisco, 1998). *J. Algorithms*, **37,** 1 (2000) 218–235. ⇒80, 81